\documentclass[aip,apl,reprint,groupedaddress,amsmath,amssymb]{revtex4-1}

\usepackage{graphicx}

\begin{document}


\title{Electron Spin Resonance with up to 20 Spin Sensitivity Measured using a Superconducting Flux Qubit}


\author{Rangga P. Budoyo}
\altaffiliation[Present Address: ]{Centre for Quantum Technologies, National University of Singapore, Singapore}
\email{cqtrpb@nus.edu.sg}
\affiliation{NTT Basic Research Laboratories, NTT Corporation, 3-1 Morinosato Wakamiya, Atsugi, Kanagawa 243-0198 Japan}
\author{Kosuke Kakuyanagi}
\affiliation{NTT Basic Research Laboratories, NTT Corporation, 3-1 Morinosato Wakamiya, Atsugi, Kanagawa 243-0198 Japan}
\author{Hiraku Toida}
\affiliation{NTT Basic Research Laboratories, NTT Corporation, 3-1 Morinosato Wakamiya, Atsugi, Kanagawa 243-0198 Japan}
\author{Yuichiro Matsuzaki}
\altaffiliation[Present Address: ]{National Institute of Advanced Industrial Science and Technology,  Tsukuba, Ibaraki, Japan}
\affiliation{NTT Basic Research Laboratories, NTT Corporation, 3-1 Morinosato Wakamiya, Atsugi, Kanagawa 243-0198 Japan}
\author{Shiro Saito}
\email{shiro.saito.bx@hco.ntt.co.jp}
\affiliation{NTT Basic Research Laboratories, NTT Corporation, 3-1 Morinosato Wakamiya, Atsugi, Kanagawa 243-0198 Japan}


\date{\today}

\begin{abstract}
We report on electron spin resonance spectroscopy measurements using a superconducting flux qubit with a sensing volume of 6 fl. The qubit is read out using a frequency-tunable Josephson bifurcation amplifier, which leads to an inferred measurement sensitivity of about 20 spins in a 1 s measurement. 
This sensitivity represents an order of magnitude improvement when compared with flux-qubit schemes using a dc-SQUID switching readout.
Furthermore, noise spectroscopy reveals that the sensitivity is limited by flicker ($1/f$) flux noise.

\end{abstract}


\maketitle



In recent years there has been an increased interest in using superconducting circuits to perform electron spin resonance (ESR) at millikelvin temperatures due to the improved sensitivity and the highly polarized spins. 
For example, by inductively coupling a spin ensemble with a superconducting lumped-element resonator with a small mode volume inductor that is combined with a parametric amplifier, a state-of-the-art sensitivity value of about 12 spins Hz$^{-1/2}$ has been reported~\cite{ranjan_arxiv2020a}. Sensitivity improvements utilizing quantum squeezing have also been achieved~\cite{bienfait_prx2017a}. 
Since these inductive-detection schemes typically employ fixed-frequency resonators that operate near their resonance, they are generally limited to magnetic field-dependent studies. However, to fully characterize more complicated spin systems (for example, anisotropic systems, or systems with hyperfine and quadrupole interactions), magnetic field- and frequency-dependent ESR spectroscopy is required~\cite{wiemann_apl2015a, chen_prb2018a}. This can be achieved by using a broadband waveguide~\cite{wiemann_apl2015a}, a tunable resonator~\cite{chen_prb2018a}, or by using the magnetic flux detection capabilities of superconducting loop structures such as a direct current-superconducting quantum interference device (dc-SQUID)~\cite{toida_apl2016a, yue_apl2017a}, a frequency-tunable Josephson bifurcation amplifier (JBA)~\cite{budoyo_prm2018a}, or a flux qubit~\cite{bal_natcomms2012a, toida_commphys2019a}.  Recently flux qubit ESR spectroscopy with sensitivity of 400 spins Hz$^{-1/2}$ has been reported~\cite{toida_commphys2019a}.


In this paper, we report on further improvements in the sensitivity of ESR using a flux qubit, which is achieved by changing the readout scheme. The previous flux qubit ESR spectroscopy was performed using a dc-SQUID switching readout~\cite{toida_commphys2019a}. 
However, in this measurement protocol, the repetition rate is limited by heating due to the dc-SQUID switching to a voltage state. 
Consequently the qubit readout is now replaced with a tunable JBA switching readout~\cite{lupascu_prl2006a, vijay_rsi2009a}, which allows an order of magnitude higher repetition rates as the JBA remains in a superconducting state thus creating no heat on the sample chip. In addition a flux qubit with a significantly smaller loop size is employed hence reducing the sensing volume by a factor of 8. 
This modification yields a sensitivity improvement by an order of magnitude to 20 spins for a 1 second measurement, making this result comparable to the best sensitivity values for inductive-coupling schemes employing a fixed-frequency resonator~\cite{probst_apl2017a}. Lastly, noise spectroscopy reveals that this sensitivity is limited by $1/f$ flux noise in the SQUID structure of the JBA.

\begin{figure}[tpb]
\includegraphics[width=0.95\columnwidth]{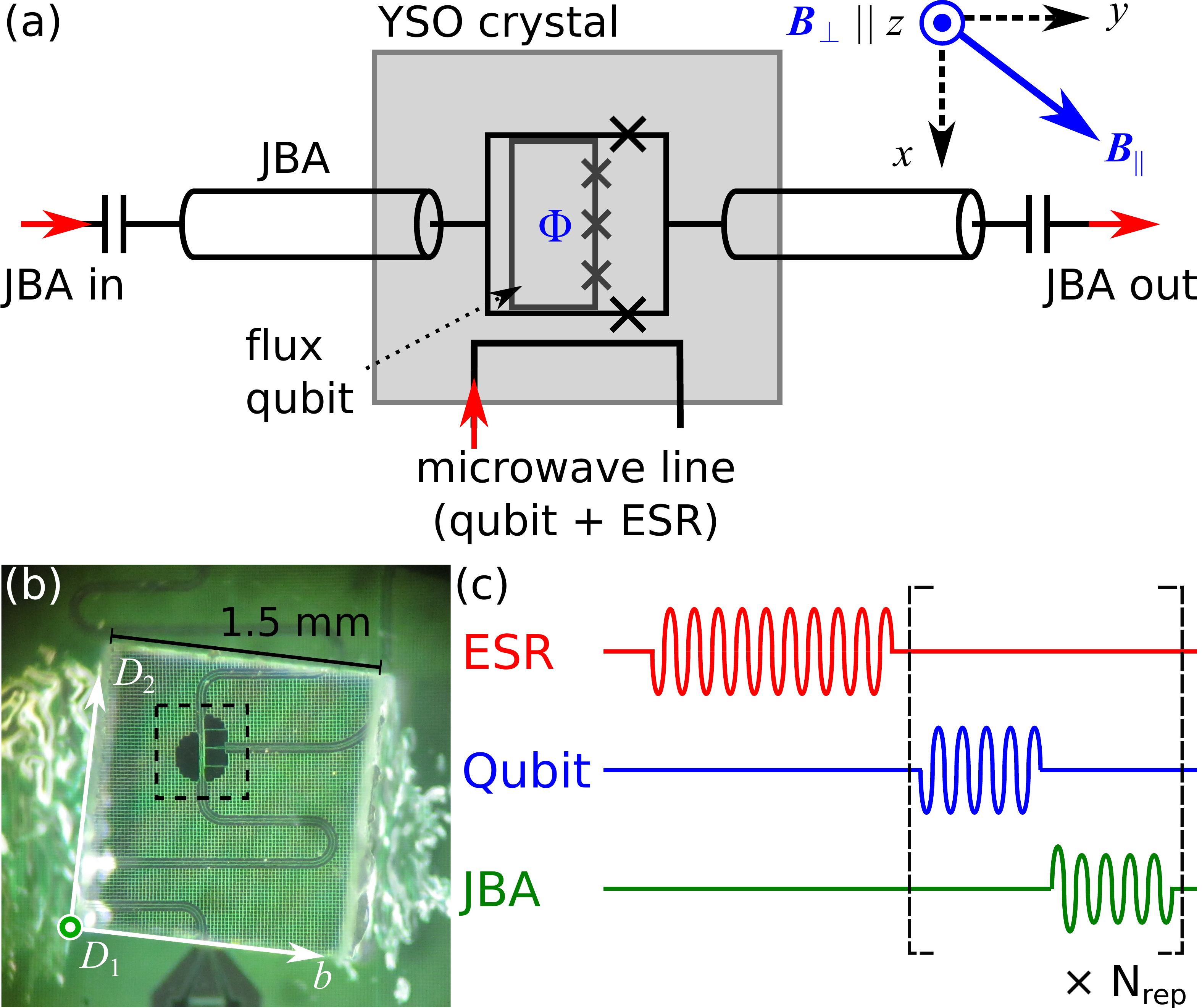}%
\caption{\label{fig:fig1combined} (a)~A schematic diagram of the ESR spectroscopy executed using a flux qubit that is readout with a JBA. (b)~A photograph of $^{167}$Er:YSO crystal mounted on top of the flux qubit chip. The white arrows correspond to the YSO crystal axes. The black dashed square marks the location of the flux qubit. (c)~The pulse sequence utilized for the ESR measurements.}
\end{figure}

Figure~\ref{fig:fig1combined}(a) shows a simplified measurement schematic. A standard three junction flux qubit with a 6 $\mu$m$^2$ loop size is embedded inside and shares its edges with a dc-SQUID of the JBA~\cite{vijay_rsi2009a, kakuyanagi_njp2015a}. A single microwave line was used for both the qubit and the ESR excitation signals. 
An isotopically purified Er$^{3+}$ doped yttrium orthosilicate (Y$_2$SiO$_5$/YSO) crystal (91.8\% $^{167}$Er isotopes with electron spin $S=1/2$ and nuclear spin $I=7/2$, with 50 ppm total Er abundance) was used in this study and it was placed on top of the qubit chip as shown in  Fig.~\ref{fig:fig1combined}(b). 
Er dopants can lie on two crystallographic sites in YSO crystal, with two subclasses for each site related by $C_2$ rotation.
A 2d vector magnet enabled magnetic field $\vec{B}_\parallel$ to be applied parallel to the surface of the chip to polarize the spin ensemble. A second, much smaller magnetic field $\vec{B}_\perp$ was applied perpendicular to the chip to flux bias the qubit. The system was then cooled to millikelvin temperature ranges using a dilution refrigerator, that was connected to a standard flux qubit measurement setup.

The system was initially characterized by measuring the magnetization induced by the spin ensemble on the qubit at different temperature $T$ and $B_\parallel$ values. More specifically $\vec{B}_\parallel$ was oriented along the $D_2$ axis of the crystal, where the two subclasses for each site are equivalent.
Due to the anisotropy of the $g$-factor tensor of Er:YSO, there is a nonzero magnetization perpendicular to the chip surface. 
For each $T$ and $B_\parallel$ value, the qubit spectroscopy was performed at a range of applied flux bias $\Phi$ values and  the resultant spectra were fitted for the qubit frequency $f_q$ given by the standard flux qubit expression
$hf_q(\Phi) = \sqrt{\Delta^2 + [\varepsilon(\Phi)]^2}$, where $\Delta$ is the flux qubit energy gap. The flux-dependent detuning is given by $\varepsilon(\Phi) = 2I_p  [\Phi - \Phi_\text{off}(B_\parallel,T)-\Phi_0/2]$, where $\Phi_0$ is the flux quantum and $I_p$ is the persistent current ($I_p\approx330$ nA for the qubit used). The offset flux $\Phi_\text{off}$ depends on $T$ and $B_\parallel$ and it includes the effect of the magnetization from the spins, any stray magnetic field from the environment, any misalignment of $B_\parallel$, and the possible presence of vortices. 
Since $\Phi_\text{off}$ is measured after a change in $T$ while $B_\parallel$ is unchanged, the latter three effects are identical with the same $B_\parallel$ and different $T$. As a result, the change in magnetization flux due to temperature change  can be defined as $\delta\Phi_m(B_\parallel, T_1, T_2) = \Phi_\text{off}(B_\parallel,T_2) - \Phi_\text{off}(B_\parallel,T_1)$.

\begin{figure}[tpb]
\includegraphics[width=\columnwidth]{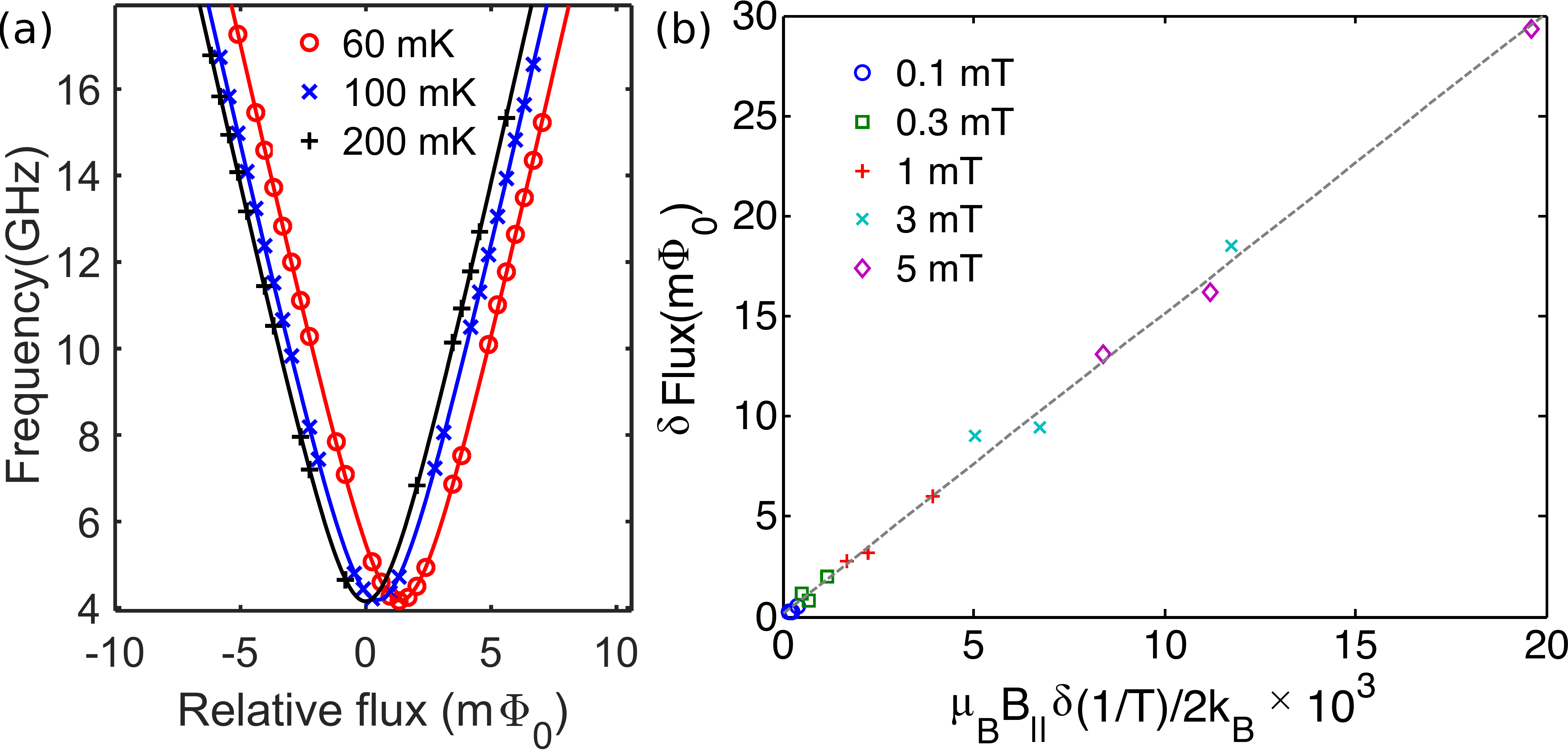}%
\caption{\label{fig:spindet} (a)~The flux qubit transition frequency as a function of flux bias at different temperatures with $\vec{B}_{\parallel}=0$ (note the relative flux is referenced to the frequency minimum for the data acquired at $T=200$ mK). The symbols are the data and the solid lines are fit detailed in the main text. (b)~The change in flux detected by the qubit at different magnetic fields and temperatures minus the shift detected at  $\vec{B}_{\parallel}=0$ shown in (a). $B_\parallel$ ranges from 0.1 to 5 mT and $T$ ranges from 50 to 200 mK. The symbols correspond to data at different fields, and the dashed line is a linear fit that is also detailed in the main text.}
\end{figure}

Figure~\ref{fig:spindet}(a) shows a plot of $f_q$ as a function of 
$\Phi$ with $B_\parallel=0$ for several temperatures, where an apparent change in magnetization is observed with a change in temperature.
Although no spin magnetization is expected when total $\vec{B}=0$, a small magnetization may emerge from the applied nonzero $B_\perp$, and/or from a stray environmental magnetic field. 
To account for these effects, $\delta\Phi_m(B_\parallel, T_1,T_2) - \delta\Phi_m(0, T_1,T_2)$ is plotted as a function of $\mu_B B_\parallel \delta(1/T)/2k_B$ in Fig.~\ref{fig:spindet}(b), with $\mu_B$ the Bohr magneton, and $\delta(1/T) = 1/T_2 - 1/T_1$.

For a spin-1/2 system with zero nuclear spin, the spin magnetization follows 
\begin{equation}
\Phi_m \: = \: \Phi_s \,\tanh\left(\frac{\mu_B g B_\parallel}{2 k_B  T}\right),
\label{eq:spindet_orig}
\end{equation}
 where $\Phi_s$ is the saturation magnetization level and $g$ is the g-factor. 
 For $\mu_B g B_\parallel \ll 2 k_B T$,
 \begin{equation}
\Phi_m \: \approx \: \Phi_s \,\frac{\mu_B g B_\parallel}{2 k_B  T}.
\label{eq:spindet}
\end{equation}
For $^{167}$Er:YSO contributions from the two sites, the anisotropy of $g$-factor tensors, as well as the presence of nuclear spins ($I=7/2$) need to be considered, but simulations~\cite{toida_commphys2019a,stoll_jmr2006a} show that the magnetization for small $B_\parallel$ still approximately follows Eq.~\ref{eq:spindet} with  an effective g-factor $\tilde{g}\approx 5.2$ for this measurement configuration. The linear fit in Fig.~\ref{fig:spindet}(b) confirms that the spin magnetization follows Eq.~\ref{eq:spindet} with $\Phi_s\approx0.29\Phi_0$, and this is consistent with previous spin detection measurements of Er:YSO \cite{toida_apl2016a, budoyo_prm2018a, toida_commphys2019a}.

\begin{figure}[tpb]
\includegraphics[width=0.95\columnwidth]{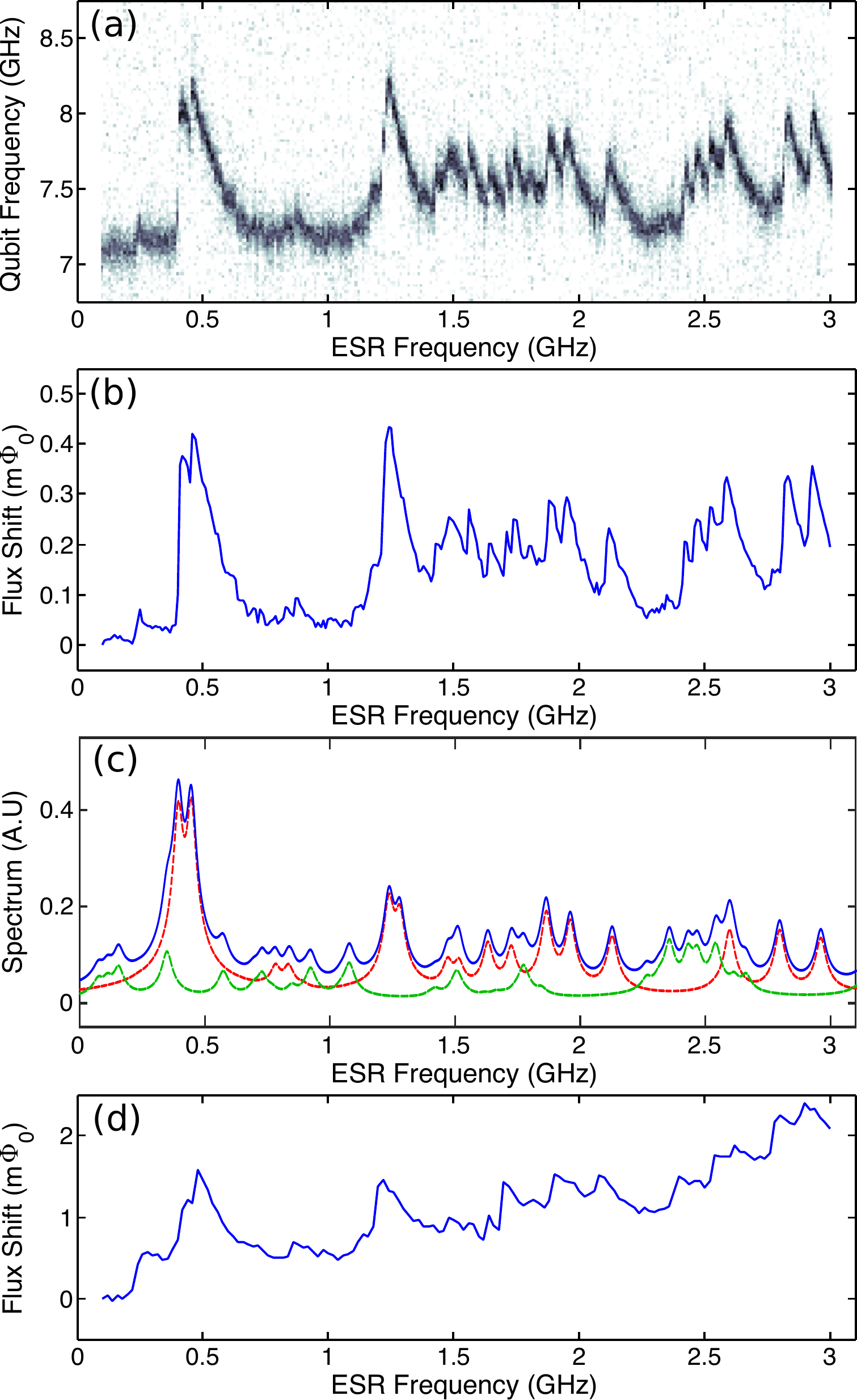}%
\caption{\label{fig:esrexample} (a)~The flux qubit spectrum as a function of ESR excitation frequency at $|\vec{B}_{\parallel}|=1.7$ mT with $T= 200$ mK. (b)~The ESR spectrum extracted from the flux qubit spectrum. 
(c)~Simulated ESR spectrum for the same magnetic field and temperature. The red (green) dashed line corresponds to the site 1 (2) spectrum, respectively, and the blue solid line corresponds to the combined spectrum.
(d)~The ESR spectrum measured using JBA ESR spectroscopy with the same magnetic field and temperature.}
\end{figure}

Next, ESR was performed by measuring the change in the qubit spectrum as a function of ESR excitation frequency. Figure~\ref{fig:fig1combined}(c) shows the pulse sequence utilized for the microwave excitations, with the applied flux $\Phi$ set at a constant value away from the qubit degeneracy point.
Figures~\ref{fig:esrexample}(a) and \ref{fig:esrexample}(b) show the ESR measurement data acquired at $T=200$ mK, with $B_\parallel=1.7$ mT oriented along the $D_2$ axis of the YSO crystal. This ESR spectrum shows asymmetric peaks due to the slowed phonon-bottleneck relaxation of Er~\cite{budoyo_apex2018a}.
This spectrum should also follow the standard spin Hamiltonian~\cite{abragam_epr1970}
\begin{equation}
{\cal H} \, = \, \mu_B \vec{B}\cdot\boldsymbol{g}\cdot\boldsymbol{S} +   \boldsymbol{I}\cdot\boldsymbol{A}\cdot\boldsymbol{S} +   \boldsymbol{I}\cdot\boldsymbol{Q}\cdot\boldsymbol{I} - \mu_n g_n   \vec{B}\cdot\boldsymbol{I},
\label{eq:spin_hamiltonian}
\end{equation}
where $\vec{B} = \vec{B}_\parallel + \vec{B}_\perp\approx \vec{B}_\parallel$, $\mu_n$  is the nuclear magneton, $\boldsymbol{S}$ is the electron spin operator, $\boldsymbol{I}$ is the nuclear spin operator, $\boldsymbol{g}$ is the electron g-factor tensor, $\boldsymbol{A}$ is the hyperfine tensor, $\boldsymbol{Q}$ is the quadrupole tensor, and $g_n$ is the nuclear g-factor. Moreover, the two crystallographic sites have different sets of anisotropic $\boldsymbol{g}$, $\boldsymbol{A}$, and $\boldsymbol{Q}$ tensors. 
Due to these characteristics, $^{167}$Er:YSO is a prime example of materials that requires frequency- and field-dependent ESR spectroscopy for a full characterization of the spin properties. In fact, recent measurements reported slightly different $^{167}$Er:YSO $\boldsymbol{g}$, $\boldsymbol{A}$, and $\boldsymbol{Q}$ values \cite{guillotnoel_prb2006a, chen_prb2018a, horvath_prl2019a} depending on the measurement methods, microwave frequencies, and magnetic field directions and strengths used. 
Accurate values of these spin tensors are of interest due to the predicted existence of zero first-order Zeeman (ZEFOZ) transitions with significantly improved coherence \cite{mcauslan_pra2012a}.
%
Fig.~\ref{fig:esrexample}(c) shows the simulated ESR spectrum \cite{stoll_jmr2006a} of $^{167}$Er:YSO for the same field value as in Figs.~\ref{fig:esrexample}(a). The simulation used the most recently reported spin tensor values, obtained using Raman heterodyne spectroscopy for site 1 \cite{horvath_prl2019a} and using a tunable cavity ESR spectroscopy for site 2 \cite{chen_prb2018a}. The simulated transition frequencies and strengths show good agreement with the spectrum taken using the flux qubit. Flux qubit spectroscopy in various magnetic field orientation and strengths should contribute further in the determination of $^{167}$Er:YSO spin tensors.

%
%



\begin{figure}[tpb]
\includegraphics[width=0.8\columnwidth]{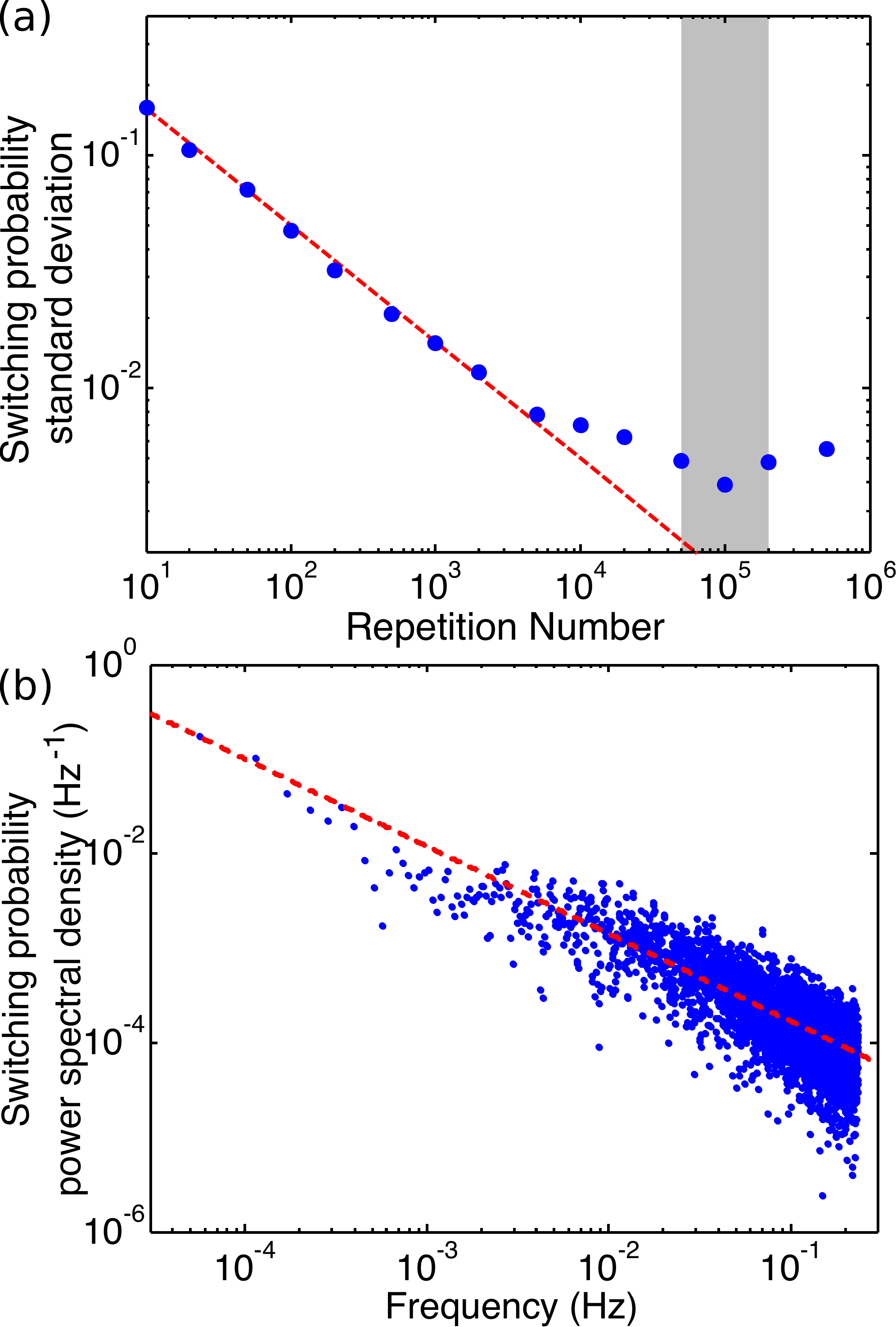}%
\caption{\label{fig:Fig5combined} (a)~The standard deviation in the qubit switching probability as a function of repetition number with the qubit at 60 mK and $B_\parallel=0$. The symbols are the experimental data and the dashed line is a calculated binomial standard deviation. The shaded region corresponds to the number of repetitions within 1 second for a typical experimental repetition time of 5 to 20 $\mu$s. (b)~The Welch spectral density of the switching probability that is obtained from repeated measurement of $P_\text{sw}$ over 7 hours. The dashed line shows a fit to the $1/f$ noise model.  }
\end{figure}

The flux qubit has an area of $6$ $\mu$m$^2$ with an effective thickness~\cite{marcos_prl2010a, toida_apl2016a} of $\sim1$ $\mu$m which yields an effective sensing volume of $\sim6$ fl and thus $N_v\approx6\times10^6$ spins within this volume. The optimal sensitivity  of the flux qubit ESR scheme (see also the Supplementary Material) can be expressed as \cite{toida_commphys2019a}
\begin{equation}
N_\text{min} \:  = \: \delta P_\text{sw} \left|\frac{\partial f_q}{\partial P_\text{sw}}\right|\frac{\delta N}{\delta f_q} 
\: = \: \delta P_\text{sw} \frac{4h\gamma_q}{3\sqrt{3}VI_p}\frac{\delta N}{\delta\Phi},
\label{eq:sensitivity}
\end{equation}
where $\delta P_\text{sw}$ is the JBA switching probability noise, $V$ is the measurement visibility, $\gamma_q$ is the line width of the qubit spectrum, $\delta\Phi/\delta N$ is the magnetic flux generated by a single spin, and $\delta f_q/\delta N$ is the change in the flux qubit's frequency from the spin magnetization.
In the standard flux qubit spectroscopy scheme, optimal values for $V\approx0.23$ and $\gamma_q\approx32$ MHz at temperatures below 100 mK. 
The spin polarization detection measurement enables $\delta N / \delta\Phi = N_v/\Phi_s$ to be derived.
To estimate $\delta P_\text{sw}$, several hundred sequential measurements of switching probability $P_\text{sw}$ were performed with varying repetition numbers $N_\text{rep}$ at similar bias conditions to those used in the ESR spectroscopy where $P_\text{sw}\approx0.5$. Figure~\ref{fig:Fig5combined}(a) shows the resulting switching probability standard deviation $\sigma P_\text{sw}$ as a function of $N_\text{rep}$ with a repetition time of 10 $\mu$s.
For low $N_\text{rep}$ values, the standard deviation follows the binomial standard deviation $\sqrt{P_\text{sw}(1-P_\text{sw})/N_\text{rep}}$ that approaches a constant value of $5\times10^{-3}$ at large $N_\text{rep}\gtrsim5000$. 
Assuming $\delta P_\text{sw} = 2 \sigma P_\text{sw}$, a typical repetition time between 5 and 20 $\mu$s infers a sensitivity of about 20 spins in a 1 second measurement.

This performance represents nearly three orders of magnitude improvement than the $10^4$ spin sensitivity for the JBA ESR \cite{budoyo_prm2018a}. Direct comparison between the flux qubit ESR and JBA ESR can be made by comparing Fig.~\ref{fig:esrexample}(b) and Fig.~\ref{fig:esrexample}(d), which is the ESR spectrum of the same Er:YSO sample obtained using the same JBA taken at the same magnetic field and temperature.
Both spectra show the same general structure, but the flux qubit-obtained spectrum can resolve more spin transitions as would be expected from a more sensitive measurement scheme.
More significantly, the 20 spin sensitivity value also corresponds to an order of magnitude improvement than the sensitivity determined from the flux qubit ESR using the dc-SQUID switching readout \cite{toida_commphys2019a}.
Since the repetition rate in the current measurement is about 20 times faster than the repetition rate using the dc-SQUID, a factor $\sqrt{20}\approx4.5$ improvement in sensitivity is expected by changing the readout scheme. 
The other sensitivity improvements arose from the greater visibility with the JBA readout in comparison to dc-SQUID switching readout, as well as the qubit's narrower line width.

To understand the limiting factor for $\delta P_\text{sw}$ at large $N_\text{rep}$, a continuous measurement of $P_\text{sw}$ was performed for approximately 7 hours. Figure~\ref{fig:Fig5combined}(b) shows the Welch power spectral density calculated from the resultant data where slow drifts due to any changes in the measurement environment (effective spin temperature or ambient field) are accounted for. This noise spectrum shows a $1/f$ (flicker) noise behavior given by
\begin{equation}
S_{P_\text{sw}}(f) \: = \: \frac{A_{P_\text{sw}}}{(f/\text{1 Hz})^\alpha},
\label{eq:spec_density}
\end{equation}
where $A_{P_\text{sw}}$ is the noise level at 1 Hz with a best fit value of $\alpha\approx0.93$ [red dashed line in Fig.~\ref{fig:Fig5combined}(b)]. Flicker noise behavior is commonly seen in the flux noise of SQUIDs \cite{wellstood_apl1987a, anton_prl2013a, kumar_prappl2016a} and  superconducting qubits \cite{yoshihara_prl2006a, bialczak_prl2007a, bylander_natphys2011a} with a typical $\alpha$ of  0.9, and a 1 Hz noise level of $A_\Phi\sim(1\mu\Phi_0)^2$/Hz. 
The presence of flicker noise limits the minimum detectable signal \cite{mcdowell_optexp2008a}  in a sensor and thus causes saturation of $\delta P_\text{sw}$ in Fig.~\ref{fig:Fig5combined}(a). 
Assuming $\delta P_\text{sw}$ originates from flux noise, the flicker flux noise level in the JBA SQUID loop can be estimated as $A_\Phi = A_{P_\text{sw}} (\partial \Phi / \partial P_\text{sw})^2 \approx (5\mu\Phi_0)^2$/Hz. 
Since this noise level is consistent with previously reported values for SQUIDs and flux qubits, it can be surmised that the ESR sensitivity of this setup is limited by the flux noise from the SQUID. 
The qubit's flux noise can also be independently measured using Ramsey interferometry \cite{yan_prb2012a} and dissipation spectral analysis methods \cite{bylander_natphys2011a, yan_natcomms2013a}, with similar noise levels. As qubit flux noise contributes to $\gamma_q$ in Eq.~\ref{eq:sensitivity}\cite{yoshihara_prl2006a, bylander_natphys2011a}, it also limits the sensitivity.

Although the sensitivity is comparable to the best values for superconducting devices, several approaches to further improve the sensitivity can be identified. For instance, surface spins are thought to contribute to $1/f$ flux noise~\cite{sendelbach_prl2008a,kumar_prappl2016a, degraaf_prl2017a} and surface treatments have been shown to reduce their $1/f$ noise level in some SQUIDs~\cite{kumar_prappl2016a}.  
Furthermore, the sensitivity quantified in Eq.~\ref{eq:sensitivity} depends on the noise level in the switching probability and also $\partial \Phi / \partial P_\text{sw}$, which can be reduced by optimizing the JBA design and bias level. 
Longer-lived flux qubits \cite{stern_prl2014a, yan_natcomms2016a, abdurakhimov_apl2019a} have reduced $\gamma_q$ but also smaller persistent current $I_p$. Thus, it remains as a future work whether such qubits improves the sensitivity.
Additionally, implementation of alternate flux qubit readout schemes such as the commonly used dispersive readout with parametric amplification \cite{lin_apl2013a} can be explored for possible sensitivity improvements.

In summary, an improvement in the sensitivity of ESR is demonstrated using a flux qubit that is readout via a Josephson bifurcation amplifier which enables the detection of an estimated 20 spins in 1 second measurement. More fundamentally this sensitivity is found to be limited by the intrinsic $1/f$ flux noise of the dc-SQUID loop of the JBA. 


See the Supplementary Material for the comparison between the ESR signal detected by flux qubit with the ESR signal detected using a conventional cavity ESR spectrometer, and their dependences to the relaxation time $T_1$ and coherence time $T_2$ of the detected spins.

The authors thank I. Mahboob for valuable discussions and critical reading of the manuscript and W. J. Munro and Y. S. Yap for valuable discussions. This work was supported by CREST (JPMJCR1774), JST.

\bibliography{FQESR}

\end{document}